\documentclass{article}
\usepackage{spconf,amsmath,graphicx}

\usepackage{xcolor}
\usepackage[switch]{lineno}
\usepackage{cite}
\usepackage{url}
\usepackage{lipsum}
\usepackage{graphicx}

\usepackage{amsfonts}
\usepackage{amsmath}
\usepackage[ruled,noline,linesnumbered,noend]{algorithm2e}
\usepackage{booktabs}
\usepackage[bookmarks=false]{hyperref}

\usepackage[english]{babel} 
\usepackage{breqn}

\newcommand{\n}[1]{\mathrm{#1}} 

\usepackage{amsmath}
\usepackage{xparse}
\ExplSyntaxOn
\NewDocumentCommand{\Rowvec}{ O{,} m }
 {
  \vector_main:nnnn { p } { & } { #1 } { #2 }
 }
\NewDocumentCommand{\Colvec}{ O{,} m }
 {
  \vector_main:nnnn { p } { \\ } { #1 } { #2 }
 }

\seq_new:N \l__vector_arg_seq
\cs_new_protected:Npn \vector_main:nnnn #1 #2 #3 #4
 {
  \seq_set_split:Nnn \l__vector_arg_seq { #3 } { #4 }
  \begin{#1matrix}
  \seq_use:Nnnn \l__vector_arg_seq { #2 } { #2 } { #2 }
  \end{#1matrix}
 }
\ExplSyntaxOff

\title{Hyperspectral Neutron CT with Material Decomposition}
\name{Thilo Balke$^{1,2}$\quad Alexander M. Long$^2$\quad Sven C. Vogel$^2$\quad Brendt Wohlberg$^2$\quad Charles A. Bouman$^1$}
\address{$^1$School of Electrical and Computer Engineering, Purdue University, West Lafayette, IN 47907, USA\\
$^2$Los Alamos National Laboratory, Los Alamos, NM 87545, USA}

\begin{document}
%
\maketitle
%

\setlength{\textfloatsep}{0.4\baselineskip plus 0.01\baselineskip minus 0.01\baselineskip} 

\setlength{\abovedisplayskip}{4pt}
\setlength{\belowdisplayskip}{4pt}
\setlength{\abovedisplayshortskip}{4pt}
\setlength{\belowdisplayshortskip}{4pt}

\begin{abstract}
Energy resolved neutron imaging (ERNI) is an advanced neutron radiography technique capable of non-destructively extracting spatial isotopic information within a given material. Energy-dependent radiography image sequences can be created by utilizing neutron time-of-flight techniques. In combination with uniquely characteristic isotopic neutron cross-section spectra, isotopic areal densities can be determined on a per-pixel basis, thus resulting in a set of areal density images for each isotope present in the sample. By preforming ERNI measurements over several rotational views, an isotope decomposed 3D computed tomography is possible. 

We demonstrate a method involving a robust and automated background estimation based on a linear programming formulation. The extremely high noise due to low count measurements is overcome using a sparse coding approach. It allows for a significant computation time improvement, from weeks to a few hours compared to existing neutron evaluation tools, enabling at the present stage a semi-quantitative, user-friendly routine application. 
\end{abstract}
%

\section{Introduction}
\label{sec:intro}

Since neutrons interact with the nucleus, rather than the electron shell such as X-rays, protons, or electrons, neutron radiography can offer complementary information to more conventional radiography probes. With neutron radiography, distinctive attenuation cross-sections, that depend both on the isotopic compositions and incoming neutron energies, can result in contrast mechanisms and material penetrabilities that are fundamentally different from x-rays~\cite{schillebeeckx2012determination}. Moreover, with the advent of short-pulsed spallation neutron sources capable of producing intense and wide neutron spectra, and ultra-fast, pixel dense neutron detectors coming to market~\cite{tremsin2009detection, TREMSIN2011415}, neutron radiographs can be further resolved based on the energy of the incoming neutrons using neutron time-of-flight (TOF) techniques~\cite{LEHMANN2009429}. This advanced technique, known as energy resolved neutron imaging (ERNI), has been developed at several neutron facilities around the world, such as ISIS~\cite{thomason2019isis}, LANSCE~\cite{lisowski2006alamos}, and J-PARC~\cite{ikeda2005current}, with broad applications ranging from nuclear fuel development~\cite{tremsin2013non} to gas pressure measurements~\cite{tremsin2017non} to 2D temperature mapping~\cite{tremsin2015spatially}. Of particular interest is the application of ERNI with epi-thermal neutrons, where many neutron absorption resonances are observed  throughout the spectrum. The underlying isotopic areal densities can be measured by fitting known cross section spectra to the measured energy-dependent transmissions on a per-pixel basis. This technique allows for not only 2D spatial mapping of specific isotopic distributions, but also 3D reconstruction of the isotopic density distribution by performing ERNI measurements over many rotational views and subsequent computed tomograpy (CT) reconstruction. Compared to dual-energy CT~\cite{liu2009quantitative} which uses two energies, in hyperspectral CT hundreds or thousands of energies can be leveraged to infer material quantities.
\begin{figure}[t!]
\centering
\includegraphics[width=1.00\columnwidth]{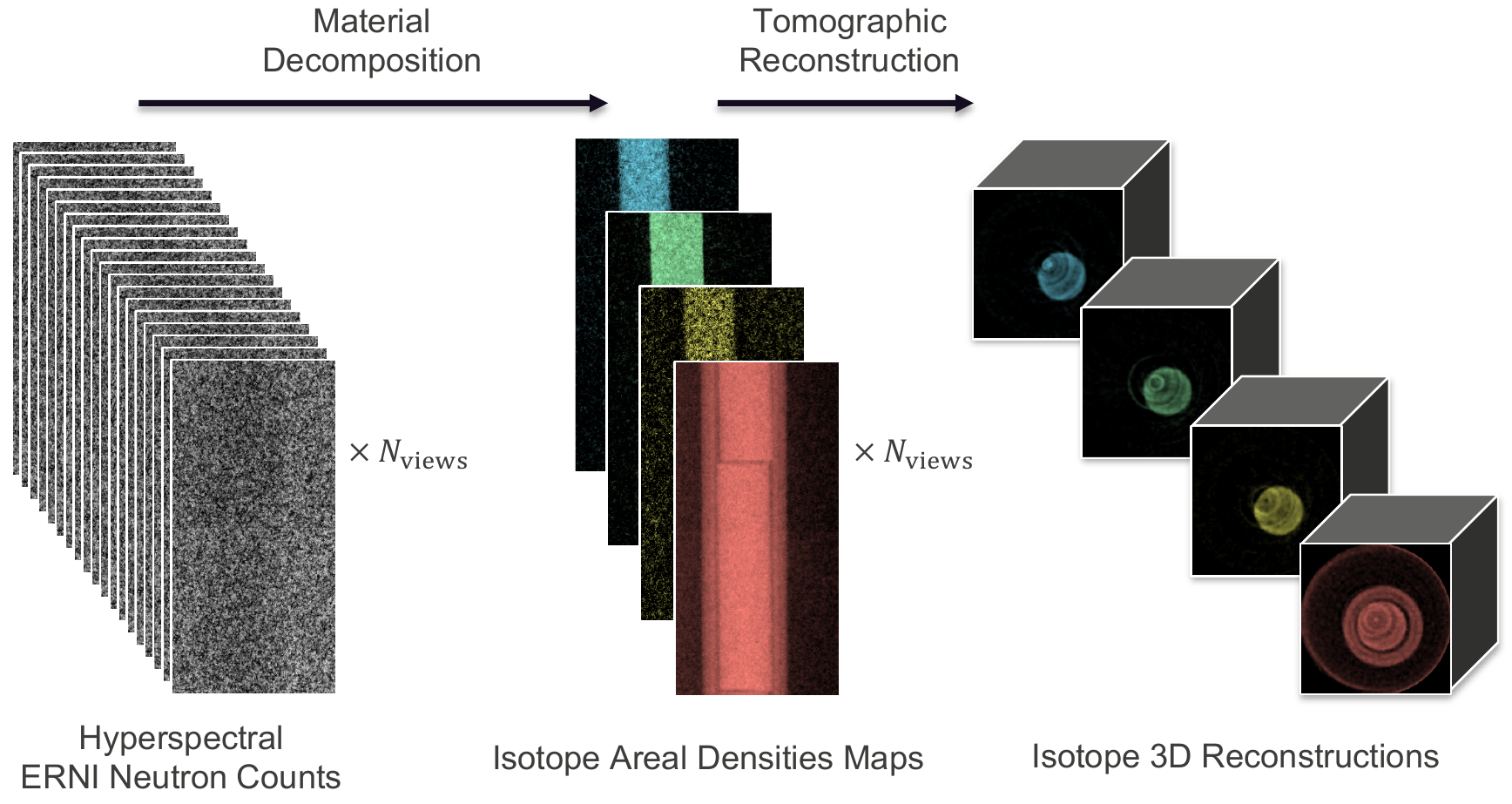}
\caption{From ERNI to material decomposed CT}
\label{fig:diagram}
\end{figure}

To date, the typical analysis process of ERNI measurements have proven to be cumbersome and computationally intensive, with the use of nuclear reaction codes, such as SAMMY~\cite{Larson2008}. With fitting routines taking seconds per transmission spectrum of a single pixel, analysis using SAMMY is computationally intensive and its application to large scale CT and radiography can require months, of computation time to render a single CT data set. Given these challenges, there is a great need for faster analysis tools that take advantage of state-of-the-art image enhancement techniques.

In this paper we present an efficient and comprehensive methodology to compute material decomposed radiographs and 3D reconstructions from ERNI radiography measurements (see Fig. \ref{fig:diagram}). We demonstrate the method on real data acquired at Flight Path 5 of the 20 Hz spallation neutron source at LANSCE~\cite{lisowski2006alamos} of the Los Alamos National Laboratory. Reliable background correction is essential for accurate computation of the transmission spectra. Although explicit formulations of the background have been proposed~\cite{schillebeeckx2012determination}, in practice this task is not performed in a standardized way~\cite{tremsin2013non}. We propose a robust, linear programming approach that is reproducible and scales computationally well with the size of the CT data set. The reconstruction of the material decomposed areal density maps is based upon a model-based iterative approach. Since ERNI measurements typically suffer from extremely poor statistics due to the flux limitations of even the world's most intense neutron sources, incorporating a Poisson counting model and Bayesian formulation in our approach helps overcome the severe noise.

\section{Reconstruction Model}
\label{sec:model}

In this section, we derive the basic equations of the physical measurements taken in the hyperspectral tomographic system. In short, we first compute a material decomposition from the hyperspectral measurements and then compute individual, material decomposed volumetric CT reconstructions. This has the advantage of reducing the dimensionality in a meaningful way, while reducing the effects of noise. 

\subsection{Hyperspectral Model}

Let $\alpha^o_{i,k}$ and $\alpha^b_{i,k}$ be the observed neutron counts of the object scan (with sample) and blank scan (without sample), respectively, at projection $i$ and TOF bin (or energy bin) $k$. Then, it is assumed that the object and blank counts are given by
\begin{equation}
\alpha_{i,k}^o = \lambda_{i,k}^o + \beta_{i,k}^o
\quad \mathrm{and} \quad
\alpha_{i,k}^b = \lambda_{i,k}^b + \beta_{i,k}^b \ ,
\end{equation}
where $\lambda$ denotes signal counts and $\beta$ the background counts whose estimation is described in section \ref{sec:background_est}.
With all these terms known, the attenuation, $Y$, is
\begin{equation}
Y_{i,k} 
= -\log
\left( 
\frac{\lambda_{i,k}^o}{\lambda_{i,k}^b}
\right)
= -\log
\left(
\frac{ \alpha_{i,k}^o - \beta_{i,k}^o }{ \alpha_{i,k}^b - \beta_{i,k}^b } 
\right) 
\ ,
\label{eq:attenuation}
\end{equation}
where the transmission is referred to as 
\begin{equation}
T_{i,k} = \frac{ \alpha_{i,k}^o - \beta_{i,k}^o }{ \alpha_{i,k}^b - \beta_{i,k}^b } \ .
\label{eq:trans_meas}
\end{equation}

Additionaly, the transmission from (\ref{eq:trans_meas}) relates the attenuation density of the sample through Beer's law,
\begin{equation}
T_{i,k}  =  \exp ( - \sum_{j} A_{i,j} U_{j,k} + B_{i,k}) \ ,
\label{eq:BeersLaw}
\end{equation}
where $A$ is the system matrix modeling the tomographic projection, $A_{i,j}$ corresponds to $i^{\n{th}}$ projection with the $j^{\n{th}}$ voxel, $U_{j,k}$ is the attenuation density of the $j^{\n{th}}$ voxel at the $k^{\n{th}}$ TOF bin, and $B$ is a noise matrix capturing the modified counting noise. By equating (\ref{eq:trans_meas}) and (\ref{eq:BeersLaw}) we can thus relate the measurements to the unknown and desired quantities, $U_{i,k}$,
\begin{equation}
Y = A U + B \ ,
\label{eq:LinearTomographic}
\end{equation}
where $Y$ is the matrix of energy-resolved attenuation measurements. $B$ can be modeled as white Gaussian noise but to justify this consider again (\ref{eq:attenuation}). As $\lambda_{i,k}$ and $\beta_{i,k}$ model independent counting events, we assume independent Poisson statistics. The variance of the attenuation, $Y_{i,k}$, given the  attenuation density, $U$, is approximated as
\begin{equation}
\n{var} (Y_{i,k}|U)
\approx
\frac{\alpha^o_{i,k} + \beta^o_{i,k}}{(\alpha^o_{i,k} - \beta^o_{i,k})^2}
+\frac{\alpha^b_{i,k} + \beta^b_{i,k}}{(\alpha^b_{i,k} - \beta^b_{i,k})^2} \ ,
\label{eq:variance}
\end{equation}
since for any differentiable function, $f$, and a random variable, $X$,  $\n{var}(f(X)) \approx \n{var}(X) [\partial f(u) / \partial u]^2 |_{u=X}$. Note that, as expected, the uncertainty of the attenuation explodes if the
counts get dominated by background. Let $V$ refer to the inverse variance in matrix form, where
\begin{equation}
V_{i,k}=1/\n{var} (Y_{i,k}|U) \quad \mathrm{and} \quad V_i = \n{diag}\{V_{i,*}\}
\end{equation}so that $V_i$ is the inverse covariance matrix of the $i^{\mathrm{th}}$ row of $Y$ given $U$.
Using this model, the conditional negative log-likelihood of the hyperspectral data, $Y$, can be derived to be
\begin{equation}
-\log p(Y|U) = \frac{1}{2} \sum_{i} \Vert Y_i - A_i U \Vert^2_{V_i} 
= \frac{1}{2} \Vert Y- A U \Vert^2_{\odot V} \ ,
\label{eq:HSLogLikelihood}
\end{equation}
where we define the weighted norms ${\Vert y \Vert^2_B  = \sum_{i,j} y_i \ B_{i,j} \ y_j}$ for a vector, $y$, and ${\Vert Y \Vert_{\odot B}^2 = \sum_{i,j} Y_{i,j}^2 B_{i,j}}$ for a matrix, $Y$, and where $Y_i=Y_{i,*}$ is the $i^{\n{th}}$ row of $Y$ and $A_i=A_{i,*}$ is the $i^{\n{th}}$ row of $A$. 

\subsection{Material Decomposed CT Model}

Directly optimizing the negative log-likelihood of (\ref{eq:HSLogLikelihood}) is not very practical for several reasons. It is assumed that the number of energy bins, $N_{\n e}$, is very large (several thousands) and the measurements are very noisy, thus directly solving for the hyperspectral attenuation densities, $U$, would imply computing a separate noisy CT for each energy which would make each reconstruction virtually unusable.  Instead, assuming only a small number of relavant isotopes, $N_{\n m} \ll N_{\n e}$,  with distinct cross section spectra present, the dimensionality and noise properties can be drastically improved.

More specifically, we represent the unknown attenuation density, $U$, as a composition of materials using the model
\begin{equation}
    U= X D \ ,
\end{equation}
where $X$ is the $N_{\n v}\times N_{\n m}$ matrix representing the material decomposed reconstruction, $N_{\n v}$ is the number of voxels, and $D$ is the $N_{\n m} \times N_{\n e}$ dictionary of spectral responses (cross sections) for each isotope. Similarly, the attenuation measurements are modeled as a composition of materials
\begin{equation}\vspace{-0.15cm}
    Y= Z D + B \ , 
    \label{eq:modelZ}
\end{equation}
where $Z=AX$ is the $N_{\n p}\times N_{\n m}$ matrix of areal densities and $N_{\n p}$ is the number of projections measured. 
Combining these equations, the forward model for our material decomposed tomographic data is given by

\begin{equation}
-\log p(Z|X) 
=
\frac{1}{2} \sum_{i} \Vert Z_i - A_i X  \Vert^2_{W_i}.
\label{eq:TransformedHSForwardModel}
\end{equation}
In this case, the inverse covariance matrices, $W_i$, are not necessarily diagonal and can be approximated from the original covariance matrices, $V_i$,
\begin{equation}
W_i = \mathbb{E} \left[ (BD^+)^\top (BD^+) \right] = (D^+)^\top V_i  D^+ \ ,
\label{eq:trans_wei}
\end{equation}
where ${D^+ =D^\top (DD^\top)^{-1}}$  is $N_{\n e} \times N_{\n m}$ Moore–Penrose pseudo-inverse of $D$.

This new model has the immediate benefit of reduced dimensionality, but it retains the problem that the covariance matrices, $W_i$, are not necessarily diagonal, as opposed to the $V_i$'s. However, in many cases the $W_i$ matrices can be approximated as being diagonal if the dictionary entries of the cross section matrix, $D$, are approximately orthogonal. With this assumption, one can separate the weights into $N_{\n m}$ diagonal matrices, $W^{(m)}$, where we define ${(W^{(m)})_{i,i} = (W_i)_{m,m}}$ such that we get $N_{\n m}$ individual conventional CT problems
\begin{equation}
-\log p(Z|X) 
\approx
\frac{1}{2}
\sum_m
\Vert  Z^{(m)} - A X^{(m)} \Vert_{W^{(m)}}^2 \ ,
\label{eq:separateCT}
\end{equation}
where $X^{(m)} = X_{*,m}$ and $Z^{(m)} = Z_{*,m}$ are the material decomposed reconstructions and sinograms, respectively. In this case,  each material reconstruction, $X^{(m)}$, derived from material density map, $Z^{(m)}$, can be computed independently with the use of a CT reconstruction algorithm minimizing the log-likelihood of (\ref{eq:separateCT}) for the unknown, $X^{(m)}$, in conjunction of an appropriate prior model.

The reconstruction follows a two step approach. In the first step the areal density maps, $\hat{Z}$, are estimated as described in section \ref{sec:mad}. In the second step the tomographic reconstruction is estimated using the likelihood model of~(\ref{eq:separateCT}) with the estimate, $\hat{Z}$, in the place of $Z$.

\subsection{Material Decomposition of Neutron Attenuation}
\label{sec:mad}

Before the tomographic reconstruction is performed, the material decomposed sinogram, $\hat{Z}$, is estimated from the hyperspectral attenuation measurements, $Y$, given by (\ref{eq:modelZ}), $Y = ZD + B$.
This quantity is estimated via a variant of Basis Pursuit Denoising (BPDN)~\cite{bpdn}
\begin{equation}
\hat{Z}= \arg \min_{Z \geq0} \left\lbrace 
\Vert D^\top Z^\top - Y^\top \Vert_{\odot V}^2
+\rho \Vert Z^\top \Vert_1
\right\rbrace \ ,
\label{eq:decomposition_bpdn}
\end{equation}
where $\rho>0$ is a regularization parameter that controls the strength of the $\ell_1$ regularization term that promotes a sparse solution of low noise material maps. Note that this is a custom form of BPDN with an unconventional data weighting due to the fact that $(D^\top Z^\top - Y^\top)$ and $V$ are matrices. The positivity constraint on the areal densities narrows down the search manifold to the non-negative quadrant. Since the cost function of~(\ref{eq:decomposition_bpdn}) is separable across projections, the problem can be solved independently for each pixel.

Importantly, equation~(\ref{eq:decomposition_bpdn}) dramatically reduces the dimensionality of the problem. This is because the number of columns in $Z$ is the number of materials, $N_{\mathrm{m}}$ (6 in our experiment), while the number of columns of $Y$ is the number of TOF bins, $N_{\mathrm{e}}$ (2290 in our experiment).

\subsection{Background Estimation}
\label{sec:background_est}

\begin{figure}[t!]
\centering
\includegraphics[width=0.85\columnwidth]{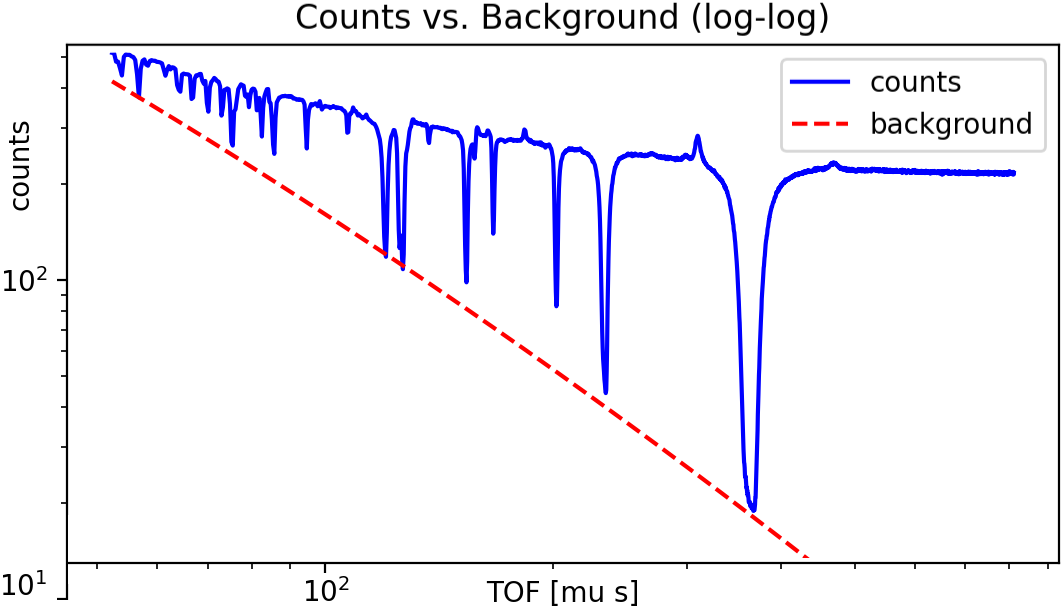}
\caption{Neutron counts over TOF with background estimate on a log-log scale. Sample is a 2.54 mm thick tantalum foil}
\label{fig:background_loglog}
\end{figure}

As (\ref{eq:attenuation}) describes, the observed neutron counts, $\alpha_{i,k}$, are modeled as a sum of direct counts, $\lambda_{i,k}$, and background counts, $\beta_{i,k}$. Fig. \ref{fig:background_loglog} shows the neutron count spectrum  with a tantalum foil sample. Due to the strong neutron absorption resonances of tantalum, deep dips in the spectrum are visible. For the deepest of those dips, the direct counts are assumed to be approximately zero. Further, when displayed in the log-log domain, as in this figure, it is assumed that the background is a smooth function within the energy region of interest. The background estimate is then intuitively the \textit{highest, smooth, lower bound of the neutron counts}. We will first describe the background correction method for a single, low noise neutron spectrum and later illustrate how to deploy this method for high noise imaging applications.

More precisely, let $\alpha = [\alpha_{i,1}, ..., \alpha_{i,N_{\n e}}]^\top$ and  $\beta = [\beta_{i,1}, ..., \beta_{i,N_{\n e}}]^\top$be the vectors corresponding to the total and background counts, respectively, while $t = [t_{i,1}, ..., t_{i,N_{\n e}}]^\top$is the vector of corresponding TOF's. In the log-log domain, we define $\tilde{\alpha} = \log \alpha$, $\tilde{\beta} = \log \beta$, and $\tilde{t} = \log t$.
The background, $\tilde{\beta}$,  is a smooth function and thus we are choosing a polynomial of low degree as a functional
\begin{equation}
\tilde{\beta} = \sum_n \tilde{t}^n x_n = G(\tilde{t})x \ ,
\end{equation}
where $x=[x_0, x_1,...]^\top$ are the polynomial coefficients and $Ax$ is the matrix formulation. To express the area under the curve in a similar way we have
\begin{equation}
\int_{\tilde{t}_1}^{\tilde{t}_N} G(\tilde{t})x \ d\tilde {t}
= \sum_n x_n
\frac{\tilde{t}_N^{i+1} - \tilde{t}_1^{i+1}}{i+1}
= - c^\top x \ .
\end{equation}
Written this way it matches the linear programming form,
\begin{equation}
\hat{x} = \arg \min_{x,\ G(\tilde{t})x \leq\tilde{\alpha}} \left\lbrace c^\top x \right\rbrace \ , 
\end{equation}
where the solution for the background is $\beta = \exp(G(\tilde{t}) \hat{x})$. 

This method solves the background estimation for a single, low noise spectrum, however, to work by itself for the whole 2D radiograph the count measurements in a TOF neutron imaging facility are usually too noisy and possibly lack the presence of totally opaque resonances. Thus, for the 2D radiographs it is assumed that a single background estimate, $\beta^*$ can be obtained from a reference measurement, $\alpha^*$, or reference region and that the background for individual projections, $i$, is proportional to the reference background
\begin{equation}
\beta_{i,*} = r_i \beta^* 
\quad \mathrm{and} \quad
r_i = \frac
{\sum_{k \in \Omega} \alpha_{i,k}}
{\sum_{k \in \Omega} \alpha_{k}^*} \ .
\end{equation}
The scale factor, $r_i$, is computed as the ratio of neutron counts in TOF regions, $\Omega$, that are assumed to be transparent to neutrons and the low noise $\alpha^*$ and $\beta^*$ are the reference total counts and background counts, respectively.

\section{Experimental Results}
\label{sec:results}

\begin{figure}[ht!]
\centering
\includegraphics[width=1.00\columnwidth]{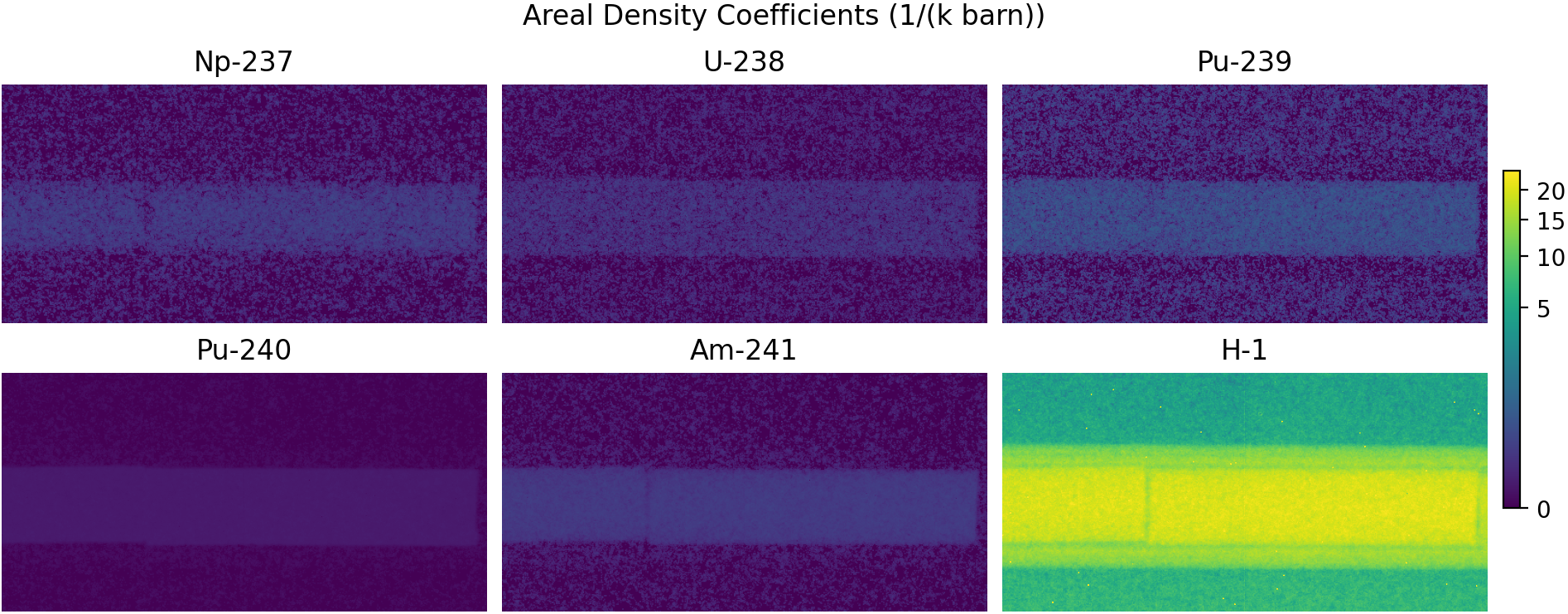}
\caption{Areal density maps ($Z$) of a single 
radiograph}
\label{fig:densityMaps}
\end{figure}

To demonstrate our approach we use ERNI measurements from LANSCE's\cite{lisowski2006alamos} neutron source, a neutron sensitive MCP detector~\cite{tremsin2009detection}, in conjunction with four Timepix readout chips with $55\!\!\times\!\!55$ $\mu$m$^2$ pitch~\cite{TREMSIN2011415}. For the CT measurement with 100 views, 2290 TOF bins each with a bin width of 320 ns was used over an energy range of approximately 1 eV to 60 keV. The sample consists of UPuZr transmutation fuel slugs, in a double-walled cylindrical steel container, with resonant isotopes in that energy region assumed to be $^{237}$Np, $^{238}$U, $^{239}$Pu, $^{240}$Pu, $^{241}$Am; and a $^{1}$H isotope placeholder capturing all  non-resonant isotopes that might be present (e.g. Zr). The material decomposition of (\ref{eq:decomposition_bpdn}), reconstructed areal density maps, $\hat{Z}$, of a single view are shown in Fig. \ref{fig:densityMaps}. The BPDN problem is optimized using an ADMM~\cite{boyd2011distributed} formulation and the SPORCO~\cite{brendt_wohlberg-proc-scipy-2017} Python package.

\begin{figure}[ht!]
\centering
\includegraphics[width=1.00\columnwidth]{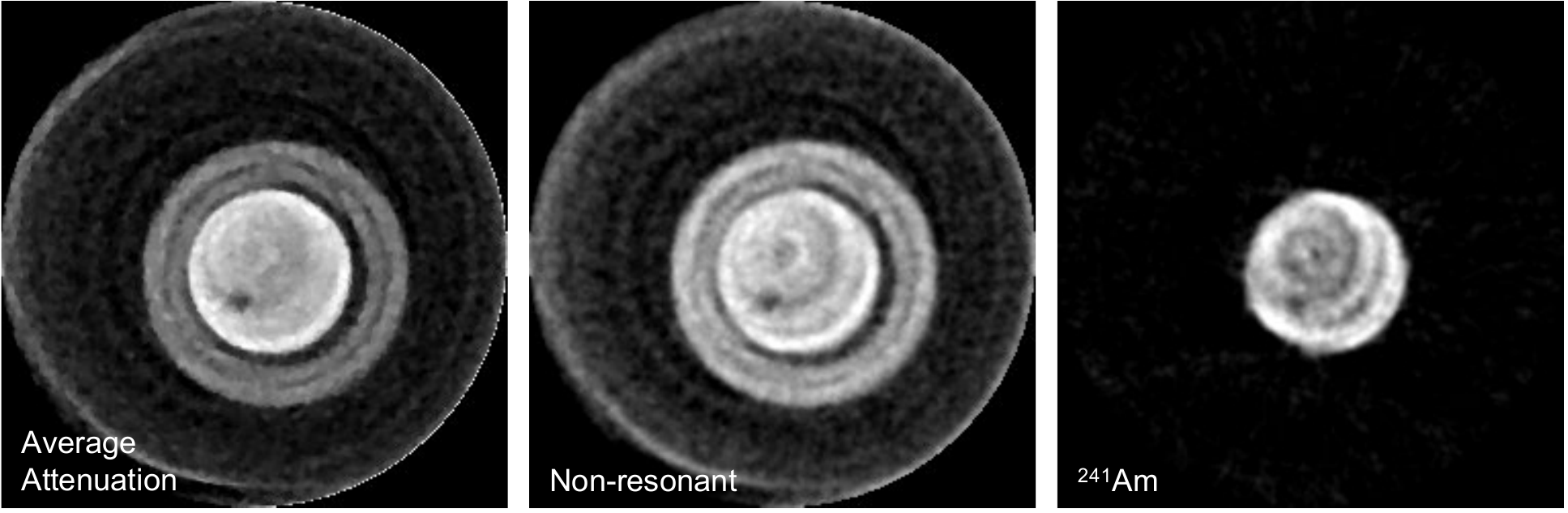}
\caption{Material decomposed CT Reconstruction ($X$) 
}
\label{fig:slices}
\end{figure}

The fast, model-based parallel beam reconstruction software, SVMBIR~\cite{wang2016HighPerformance} was used for the CT reconstruction, optimizing the log-likelihood from (\ref{eq:separateCT}) with a q-GGMRF prior model serving as regularization term~\cite{kisner2012model} for each isotope independently as in (\ref{eq:separateCT}). In addition, an average attenuation sinogram can by computed by averaging $Y$ across the energies. The CT reconstruction (axial slices) for the average attenuation, the non-resonant group, and for the $^{241}$Am isotope are shown in Fig. \ref{fig:slices}. With this reconstruction, it is clearly visible that both the fuel core and the steel container contain non-resonant isotopes, presumably Zr and Fe, whereas the $^{241}$Am only shows up in the center of the fuel slug. Also, as a void in the 7 o'clock segment off the center shows up in all isotope maps, it can be assumed that this is a void.

\begin{table}[t]
\caption{\label{tab:compTimes}Computation time comparison of the material \mbox{decomposition}. (*: interpolated, **: extrapolated)}\vspace{0.1cm}
\centering
\begin{tabular}{@{}lll@{}}
\toprule
Compuation Time & SAMMY\cite{Larson2008} & Proposed \\ \midrule
Time per pixel spectrum & 1.72 s       & 2.18 ms* \\
Time per CT data set   & 248.8 days** & 7.58 h  \\\bottomrule
\end{tabular}
\end{table}

The computation time is dominated by the material decomposition from (\ref{eq:decomposition_bpdn}) and for the data set consisting of $504 \times 248 \times 2290 \times 100$ data points (pixel, pixel, TOF bins, view angles) only approximately 7.58 hours on a 8-core 2.4 GHz machine was required. We estimate (by extrapolating from a single-spectum analysis) that it would have taken approximately 8.3 months to solve the same problem using SAMMY (see Tab.~\ref{tab:compTimes}).
\begin{table}[h!]
\caption{\label{tab:atomsPerBarn}Estimated Areal densities for a single spectrum in ${[}$atoms / 1000 barn${]}$. 
(* SAMMY did not converge using $^{1}$H) }\vspace{0.1cm}
\centering
\begin{tabular}{@{}rrr@{}}
\toprule
Areal Densities & SAMMY\cite{Larson2008}    & Proposed \\ 
\midrule
$^{237}$Np         &  0.471 & 0.577 \\
$^{238}$U          & 10.665 & 1.163 \\
$^{239}$Pu         &  1.930 & 2.284 \\
$^{240}$Pu         &  0.540 & 0.001 \\
$^{241}$Am         &  0.493 & 0.727 \\
$^{1}$H equivalent & N/A*    & 0.031 \\ \bottomrule
\end{tabular}
\end{table}
The estimated densities from a single low noise spectrum are shown in Tab.~\ref{tab:atomsPerBarn}. Although most isotopes have been estimated relatively similar to the SAMMY benchmark, both $^{238}$U and $^{240}$Pu show large deviations. This is attributed to the fact that those isotopes show very few but very intense resonances which cause the signal to reach into the noise floor. Thus, to call the proposed tool truly quantitative, an extensive comparison and evaluation is necessary, however we believe that the so far extreme speed benefits are providing a very promising tool for qualitative analysis with possibly future quantitative verification.

\clearpage

\bibliographystyle{IEEEbib}
\bibliography{strings,refs}


\end{document}